# Spin Polarization in a AlGaAs/GaAs Quantum Point Contact with in-plane side gates

N. Bhandari[1], P. P. Das[1], M. Cahay[1], R. S. Newrock[2], and S. T. Herbert[3]

[1]School of Electronics and Computing Systems, University of Cincinnati, Cincinnati, Ohio 45221, USA

[2]Physics Department, University of Cincinnati, Cincinnati, Ohio 45221, USA

[3]Department of Physics, Xavier University, Cincinnati, Ohio 45207, USA

## Abstract

We report the observation of an anomalous conductance plateau near $G = 0.5\ G_0$ ($G_0 = 2e^2/h$) in asymmetrically biased AlGaAs/GaAs quantum point contacts (QPCs), with in-plane side gates in the presence of lateral spin-orbit coupling. This is a signature of spin polarization in the narrow portion of the QPC. The appearance and evolution of the conductance anomaly has been studied at $T$=4.2K as a function of the potential asymmetry between the side gates. The observation of spontaneous spin polarization in a side-gated GaAs QPC could eventually lead to the realization of an all-electric spin-valve at tens of degrees Kelvin.

Semiconductor spintronics is one of the most promising paradigms for the development of novel devices for use in the post-CMOS era [1,2]. It is based on the simultaneous manipulation of the electron charge and spin and offers the possibility of high speed - low power devices. Many attempts have been made to achieve spin injection, detection, and manipulation by incorporating ferromagnetic materials into device architectures, with or without external magnetic fields. This results in significant design complexities. In addition, magnetic electrodes can have magnetoresistance and can also have spurious Hall voltages that can complicate device operation. The major challenge of spintronics is to avoid the use of ferromagnetic contacts or external magnetic fields and to control the creation, manipulation, and detection of spin polarized currents by purely electrical means. Some major steps towards that goal have been reported recently [3-8].

Since spin-orbit coupling (SOC) couples the electron's motion to its spin, SOC has been envisioned as a possible tool for all-electrical spin control and generation of spin-polarized currents. It has been shown that SOC can be used to modulate spin polarized currents by taking advantage of symmetry-breaking factors such as interfaces, electric fields, strain, and crystalline directions [5]. Recently, we showed that *lateral spin-orbit coupling* (LSOC) in InAs/In$_{0.52}$Al$_{0.48}$As quantum point contacts (QPCs) with in-plane side gates can be used to create a strongly spin-polarized current by purely electrical means *in the absence of any applied magnetic field* [9,10]. We studied the appearance and evolution of several anomalous conductance plateaus (in the range from 0.4 to 0.7 G$_0$ with G$_0$ = *2e$^2$/h*) in InAlAs/InAs QPCs at T=4.2K as a function of the DC offset bias $\Delta V_\text{G}$ between the two in-plane gates of the QPC [11]. We found that the number and location of the anomalous conductance plateaus strongly depend on the polarity of the DC offset bias. The anomalous plateaus appear only over an intermediate range of DC offset bias. They are quite robust and initial evidence suggests that they are highly dependent on the width of the QPC with small width giving broader plateaus [10,11]. These results were interpreted as evidence for the sensitivity of the QPC spin polarization to defects (surface roughness and impurity (dangling bond) scattering generated during the etching process that forms the QPC side walls [11]. This assertion is supported by non-equilibrium Green function (NEGF) simulations [12,13,14] of the conductance of a single QPC in the presence of dangling bonds on its walls. Our simulations show that a spin conductance polarization near 98% can be achieved despite the presence of dangling bonds and surface roughness scattering. This maximum is not necessarily reached where the conductance of the channel is equal to 0.5 $G_0$ [11].

The work described above was done at the low temperature of 4.2K. It is important to explore ways to go to higher temperatures before any practical application of a QPC spin polarizer can be envisioned. InAs, which has a high intrinsic SOC, has a short spin coherence length: about a micron at 4.2K [15]. This reduces to only tens of nanometers at ambient temperatures. This makes InAs, or any other semiconductor with a large intrinsic SOC, unsuitable for making practical devices operational at ambient temperatures. Our NEGF simulations demonstrate that a strong SOC is *not* essential to the generation of a strong spin polarization [12]. Even a very weak SOC can cause significant spin polarization provided the electron-electron (e-e) interaction is very strong. This means that QPCs made from a material like GaAs, which has a weak intrinsic SOC could also be used to generate spin polarized current by purely electrical means. GaAs has a long spin coherence length of tens of microns [16] at ambient temperatures, as compared to tens of nanometers for InAs. It is also possible to grow GaAs samples with very low electron concentration which ensures a strong e-e interaction. GaAs is a mainstream material with a mature and well-established processing technology. It also has the added advantage of a large Schottky barrier, making it relatively easy to deposit surface gates. GaAs is therefore an ideal potential candidate for developing all-electric spin devices that can be operational at temperature of a few tens of Kelvin or higher.

Here, we report the observation of a near 0.5 $G_0$ conductance plateau at T=4.2K in asymmetrically biased side-gated GaAs QPCs in the presence of LSOC. We used a two-dimensional electron gas (2DEG) formed at the hetero-interface of Si-modulation doped GaAs/AlGaAs quantum heterostructure to fabricate the QPC device (Fig.1(a)). The doped layer of AlGaAs is separated by an undoped AlGaAs layer called spacer. The thickness of this spacer layer helps control the carrier concentration in the 2DEG. Because of this spatial separation, the electrons in the 2DEG do not suffer scattering from the ionized impurities and high mobility carriers are realized. The 2DEG was characterized by Shubnikov-de Haas (SdH) and quantum Hall measurements; its carrier density and mobility were found to be $1.6 \times 10^{11}/cm^2$ and $1.9 \times 10^5$ $cm^2/VS$, respectively. Sample cleaning was performed using a procedure described in ref. [11]. A 50 nm thick polymethylcrylate (PMMA) electron beam resist was spin-coated and then exposed, using electron beam lithography, to define the narrow constriction of the QPC device. The electron dose was 65 $\mu C/cm^2$ and the accelerating voltage 10 kV. The pattern was then developed in MIBK:isopropanol (1:1) for 65s. After post-baking the sample at 115 $^0C$ for 5 min., it was etched in $H_2O:H_2O_2:H_3PO_4$ (38:1:1) for 65 sec to etch two narrow trenches (to define the QPC constriction) about 180 nm deep and 450 nm wide, as shown in Fig.1(b). Ohmic contacts were deposited using 12 nm of Ni, 20 nm of Ge and 300 nm of Au, followed by a rapid thermal annealing at 350 $^0C$ for 180s.

In the two devices reported here, the narrow portion of the QPC channel has a width (along y-direction) and length (along x-direction) around 370 nm and 400 nm, respectively, for QPC1, and around 350 nm and 400 nm, respectively, for QPC2. The electrostatic width of the QPC channel was changed by applying bias voltages to the metallic in-plane side gates, depleting the channel near the side walls of the QPC. Battery operated DC voltage sources were used to apply constant voltages $V_{G1}$ and $V_{G2}$ to the two gates. An asymmetric potential $\Delta V_G = V_{G1} - V_{G2}$ between the two gates was applied to create spin polarization in the channel. The QPC conductance was then recorded as a function of a common sweep voltage, $V_G$, applied to the two gates in addition to the potentials $V_{G1}$ and $V_{G2}$, with the current flowing in the x-direction (Fig. 1). The linear conductance $G$ ($=I/V$) of the channel was measured for different $\Delta V_G$ as a function of $V_G$, using a four-probe lock-in technique with a drive frequency of 17 Hz and a drain-source drive voltage of 100 µV. All measurements were made at $T=4.2$K. For all values of $\Delta V_G$, the gates were found to be non-leaking.

Figures 2 and 3 show the conductance of the two QPCs as a function of the sweep voltage $V_G$ for different asymmetric biases ($\Delta V_G = V_{G1} - V_{G2}$) between the gates. Because GaAs has a large surface depletion as a result of Fermi level pining by surface states [17], a large positive potential (about 12 V) was needed on both gates to obtain a conducting channel at T=4.2K. The potential on both gates was then gradually reduced in the range of a few volts making sure the channel remained open. In Figures 2 and 3, the left-most curve shows the conductance for the symmetric case, i.e., with only the common sweep voltage $V_G$ applied to the gates. The conductance curve is rather smooth, with no major features at 0.5 or 1.0 $G_0$. We attribute this to significant elastic scattering in the narrow portion of the QPC, due either to surface roughness scattering or dangling bonds at both channel/vacuum interfaces, as supported by the surface ruggedness around the QPC [10]. For the other curves, from left to right, the potential $V_{G2}$ applied to gate $G_2$ is fixed at 0 V and the potential $V_{G1}$ on gate $G_1$ is varied from 0 to 4.0 V (for QPC1) and to 4.2V (for QPC2), the latter corresponding to a large asymmetry between the two gates. As can be seen from Figures 2 and 3, an anomalous plateau (around 0.5 $G_0$) is only observed for an intermediate range of asymmetric bias $\Delta V_G$ and appears over a sweep voltage range of a fraction of a volt which is less than the 1V range observed in previously reported InAs QPCs [11]. The asymmetric bias eventually leads to spin polarization in the channel, triggered by the imbalance of the LSOC on the two sides of the channel, as discussed in detail in our earlier work with InAs QPCs [9,12]. With the increase of the asymmetric potential, the 0.5 plateau is prominent over an intermediate range of $\Delta V_G$ around 2.6 V for QPC1 and 3.3 V for QPC2 but then eventually disappears, a feature similar to the one observed with InAs QPCs [10]. When the bias asymmetry is large (with a more positive bias on gate $G_1$), electrons in the channel are squeezed towards gate $G_1$, increasing the electron concentration on that side of the channel. This leads to an increased screening of the e-e interactions near the sidewall close to gate $G_1$, quenching the onset of spin polarization in the QPC [12,14]. For QPC2, there is also a conductance anomaly around 0.3 $G_0$. QPC2 has an aspect ratio (width/length of the narrow portion of the QPC) = 1.14, slightly larger than the aspect ratio of QPC1 which is equal to 1.08. In the past, we have used NEGF simulations to show that more conductance anomalies are present in QPCs with larger aspect ratio [13]. Another possible explanation for the anomaly around 0.3 $G_0$ is the difference in the number and location of dangling bonds in the narrow portion of both QPCs which can also lead to other conductance anomalies in addition to the 0.5 structure, as supported by NEGF simulations [12].

We further confirmed (not shown here) the presence of surface scattering in our sample by measuring the magnetic field dependence of the conductance of both QPCs as a function of the sweep voltage $V_G$ for a fixed asymmetric bias between the two side gates. The magnetic field was perpendicular to the device plane or the 2DEG. As for the case of InAs QPCs reported earlier [10], it was found that, under the influence of the magnetic confinement, the 0.5 plateau evolves smoothly towards a well defined normal conductance plateau $G_0$. This indicates that magnetic confinement leads to a diminished scattering from the side walls. Transport through the channel is then near-ballistic and the normal conductance plateau is recovered.

In a preliminary effort to explore the potential of GaAs as a spintronics material for developing an all electric QPC spin polarizer, we have reported the successful observation of a near 0.5 $G_0$ plateau at T=4.2K in asymmetrically biased side-gated GaAs QPCs in the presence of LSOC. Future work will focus on improvements of the QPC design to demonstrate efficient QPC spin polarizer at higher temperature.

**Acknowledgment**

This work is supported by NSF Awards ECCS 0725404 and 1028483.

**Figure Captions**

Fig. 1: (a) The AlGaAs/GaAs heterostructure used to build the QPC device. The 2DEG is separated from the Si-doped AlGaAs layer by an undoped AlGaAs spacer layer. (b) A three-dimensional atomic force micrograph of our QPC with two in-plane gates ($G_1$ and $G_2$), fabricated using a chemical wet etching technique. The current flows in the x-direction. An asymmetric LSOC is generated using an asymmetric bias between the two gates generating an electric field in the y-direction.

Fig. 2. The conductance of QPC1 (in units of $2e^2/h$) measured as a function of the common sweep voltage $V_G$ applied to the in-plane gates, at T= 4.2 K. The sweep voltage is superimposed on potentials $V_{G1}$ and $V_{G2}$ initially applied to the gates to create an asymmetry. The left-most curve shows the conductance for the symmetric case; i.e., with only the common sweep voltage $V_G$ applied to the gates. For the other curves, from left to right, the initial potential $V_{G2}$ applied to gate $G_2$ is fixed at 0.0 V and the initial potential $V_{G1}$ on gate $G_1$ is set equal to (from left to right) 0.0, 0.5, 1.0, 1.5, 2.0, 2.5, 2.8, 3.0, 3.2, 3.4, 3.6, 3.8, and 4.0 V. These curves are shifted along the voltage axis for clarity.

Fig. 3. Similar to Fig.2 but for QPC2. The left-most curve shows the conductance for the symmetric case; i.e., with only the common sweep voltage $V_G$ applied to the gates. For the other curves, from left to right, the initial potential $V_{G2}$ applied to gate $G_2$ is fixed at 0.0 V and the initial potential $V_{G1}$ on gate $G_1$ is set equal to (from left to right) 0.0, 0.3, 0.6, 0.9, 1.2, 1.5, 1.8, 2.1, 2.4, 2.7, 3.0, 3.3, 3.6, 3.9, and 4.2V.

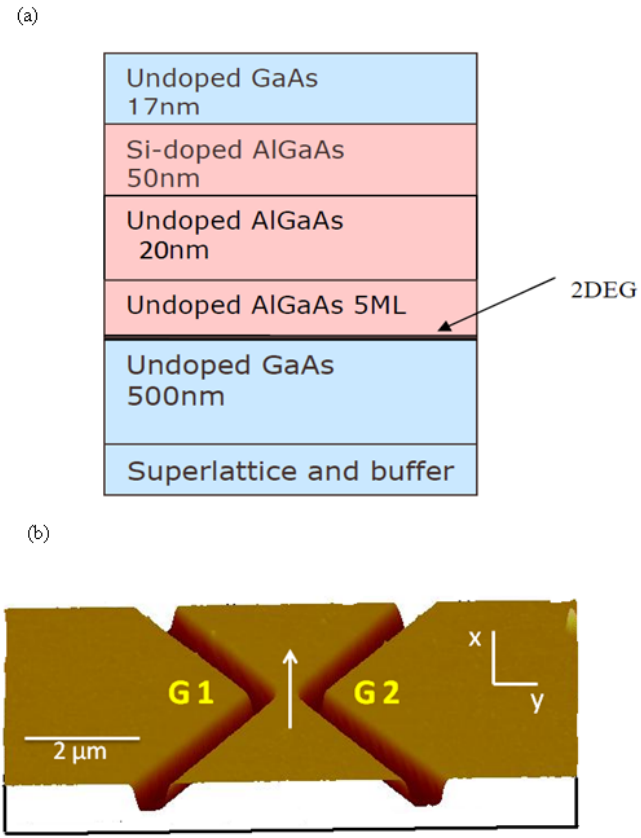

**Figure 1 (Bhandari et al.)**

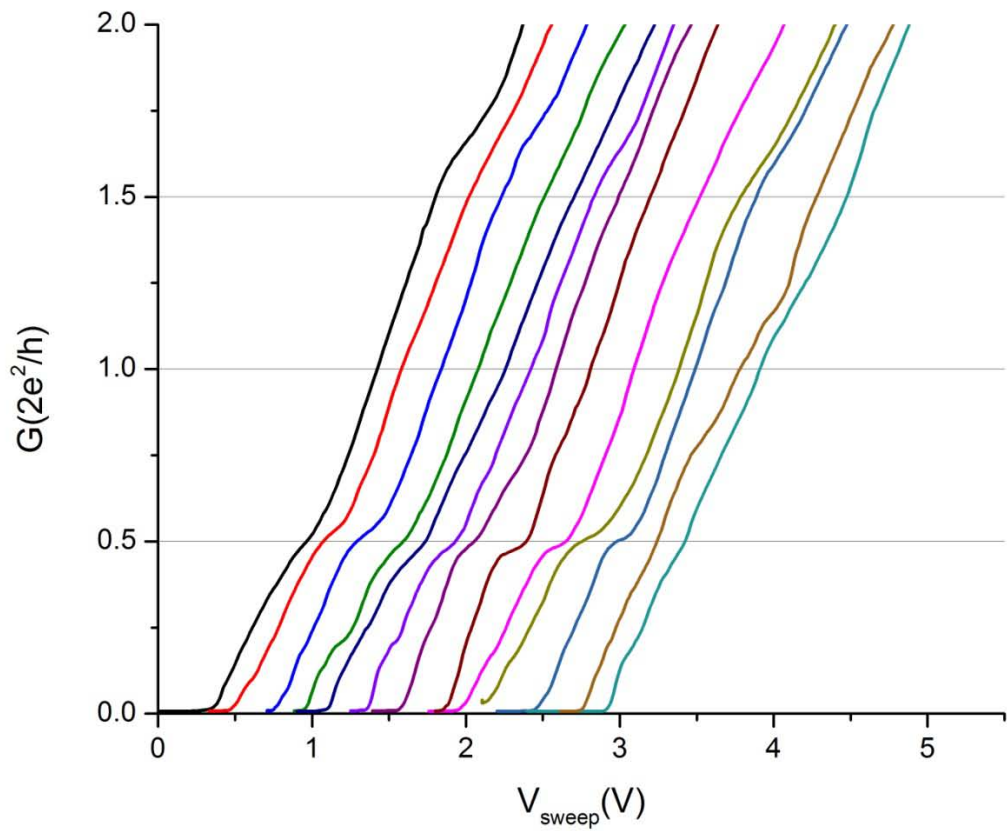

**Figure 2 (Bhandari et al.)**

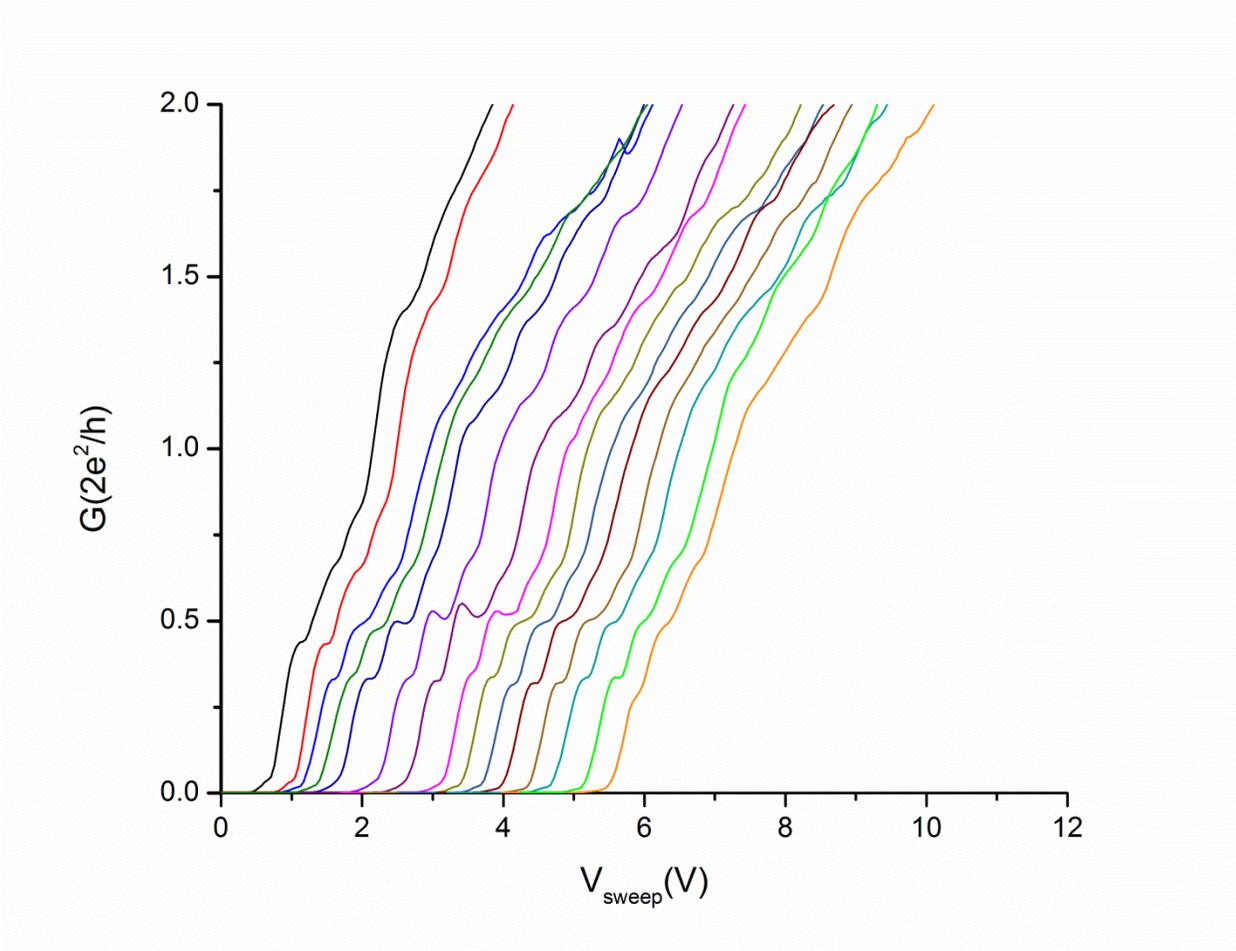

**Figure 3 (Bhandari et al.)**